# Investigation of the Privacy Concerns in AI Systems for Young Digital Citizens: A Comparative Stakeholder Analysis

Molly Campbell
*Computer Science Department*
*Vancouver Island University*
Nanaimo, Canada
molly.campbell@viu.ca

Ankur Barthwal
*Business and Management*
*Vancouver Island University*
Nanaimo, Canada
ankur.barthwal@viu.ca

Sandhya Joshi
*VIU Affiliate*
*Vancouver Island University*
Nanaimo, Canada
sandhya.joshi@viu.ca

Austin Shouli
*Computer Science Department*
*Vancouver Island University*
Nanaimo, Canada
austin.shouli@viu.ca

Ajay Kumar Shrestha
*Computer Science Department*
*Vancouver Island University*
Nanaimo, Canada
ajay.shrestha@viu.ca

*Abstract*— The integration of Artificial Intelligence (AI) systems into technologies used by young digital citizens raises significant privacy concerns. This study investigates these concerns through a comparative analysis of stakeholder perspectives. A total of 252 participants were surveyed, with the analysis focusing on 110 valid responses from parents/educators and 100 from AI professionals after data cleaning. Quantitative methods, including descriptive statistics and Partial Least Squares Structural Equation Modeling, examined five validated constructs: Data Ownership and Control, Parental Data Sharing, Perceived Risks and Benefits, Transparency and Trust, and Education and Awareness. Results showed Education and Awareness significantly influenced data ownership and risk assessment, while Data Ownership and Control strongly impacted Transparency and Trust. Transparency and Trust, along with Perceived Risks and Benefits, showed minimal influence on Parental Data Sharing, suggesting other factors may play a larger role. The study underscores the need for user-centric privacy controls, tailored transparency strategies, and targeted educational initiatives. Incorporating diverse stakeholder perspectives offers actionable insights into ethical AI design and governance, balancing innovation with robust privacy protections to foster trust in a digital age.

## I. INTRODUCTION

From recommendation algorithms to chatbots, artificial intelligence (AI) systems are increasingly becoming a part of everyday life. Young digital citizens, defined as children and young people raised in a technology-driven world, specifically engage with a wide range of AI systems on a regular basis and frequently encounter challenges related to the misuse and unauthorized access of their personal data [1]. Research shows that there are numerous privacy concerns around the use of these technologies, and young users may not be aware of these risks [2]. Therefore, it is not only important that the privacy perspectives of young digital citizens are understood, but the views of other stakeholders are also examined.

Parents, in particular, have a vested interest in the privacy policies and practices involving the technology their children use [3]. Educators also play a critical role, as they may introduce new technologies in the classroom and influence student's views by providing guidelines, perspective and context around these technological innovations [4]. By contrast, AI researchers and developers make decisions in their work that directly impact the data-collection and privacy implications of AI applications [5].

In today's increasingly interconnected world, it is important to incorporate diverse stakeholder perspectives into regulatory and governance decisions [6]. As the regulation of AI technology is being explored, it is important to consider the varying perspectives of stakeholders. Understanding their perspectives through empirical studies provides valuable context for understanding the privacy implications of AI systems and may provide insight into effective approaches for designing future regulations around AI development [7].

This paper focuses on quantitative analysis, while the qualitative insights derived from the same research are presented in a separate study [8]. To investigate privacy concerns across different stakeholder groups, three surveys were designed, one for young digital citizens, one for parents and educators, and one for AI researchers and developers. These surveys aimed to quantitatively explore and measure what the privacy concerns of each group are, by investigating five distinct validated constructs from the existing literature: Data Ownership and Control (DOC), Parental Data Sharing (PDS), Perceived Risks and Benefits (PRB), Transparency and Trust (TT), and Education and Awareness (EA). This study analyzes the responses to capture the attitudes of each stakeholder group regarding these constructs and their influence on data-sharing attitudes.

As AI systems continue to improve, it is worth considering the differing perspectives between parents/educators and AI researchers/developers. Where parents and educators play a role in determining the attitudes of young digital citizens towards AI systems, AI researchers/developers play a role in determining how these systems are designed, and what privacy concerns may arise from those design decisions [7]. A comparison of these two categories of stakeholders can help identify if parents and educators are more or less informed and/or concerned compared to researchers/developers, which is a critical indicator of their overall readiness to navigate the complex contemporary data privacy landscape.

In the next section, we present background and related works that provide the foundation for this study, including prior research on privacy concerns in AI systems and the development of the constructs. Section III describes the methodology, including the research hypotheses, survey design, and participant demographics. The results, including descriptive statistics, measurement models and Partial Least Square Structural Equation Model (PLS-SEM) employed for analysis are detailed in Section IV. Section V discusses the implications of these findings for stakeholders and the development of ethical guidelines for privacy in AI systems. Finally, Section VI concludes the paper with a summary of key contributions, limitations, and directions for future research.

## II. BACKGROUND AND RELATED WORKS

### A. Privacy Concerns in AI Systems for Young Digital Citizens

The integration of AI systems into various technologies has transformed the digital landscape, particularly for young users who engage frequently with these platforms. Young digital citizens, raised in a technology-centric environment, face unique privacy challenges that stem from their interactions with AI-powered applications. These include the collection and processing of extensive personal data, which, while enabling tailored experiences, pose risks related to data security and unauthorized access [9], [10], [11]. A study by [12] involving 1,021 caregivers and youth dads (average age 12.12 years; 42% female), found that effective parenting techniques, such as the "Educate" strategy, positively influenced youth privacy awareness, whereas less comprehensive approaches reduced concerns. This highlights how active mediation shapes youth privacy perspectives, revealing the complexity of these influences in digital interactions. Research also indicates that young users often lack the understanding necessary to comprehend how their data is used, making them susceptible to privacy breaches and exploitation [13], [14].

The implications of data misuse have been magnified by high-profile data breaches and unethical data practices on social media platforms [13], [15]. As a result, parents and educators have become more vigilant about the digital safety of young individuals, calling for greater transparency and accountability in AI systems [15]. Research, such as a study [16] that surveyed 148 younger adults (ages 18-25) and 152 older adults (ages 55+), revealed no significant age differences in privacy attitudes but highlighted a universal demand for data control and clear communication. This shows that privacy concerns are prevalent across age demographics and that robust privacy strategies are crucial for fostering trust among young digital citizens. However, the current literature indicates a gap in understanding the diverse perspectives of different stakeholders regarding these privacy issues and their potential solutions.

### B. Comparative Analysis of Stakeholder Perspectives

This study aims to bridge this gap by exploring the differing perceptions of privacy concerns between parents/educators and AI developers/researchers through various validated constructs as given in Table I. The corresponding structural equation model is shown in Fig. 1. These constructs, derived from established literature, were selected for their ability to comprehensively capture the dimensions of privacy that are critical to understanding stakeholder attitudes and identifying actionable insights for improving privacy practices in AI systems. Parents and educators play a critical role as mediators who introduce technology to young users and shape their understanding of digital safety [13], [16]. For example, [17], [18] examination of teenagers' use of AI-enabled voice assistants involved semi-structured interviews with thirty-six high school students (ages 13-15) during the COVID-19 pandemic, revealing that intrinsic motivation and privacy concerns significantly influenced technology adoption among youth. This insight emphasizes the multifaceted nature of youth interactions with AI systems and the role of social and educational influences. They are particularly interested in ensuring that AI systems provide clear communication about data usage and maintain rigorous data protection standards. Studies show that parents and educators value transparency as it directly impacts their trust in the technology and influences their comfort level in promoting its use among young people [15], [19].

In contrast, AI developers and researchers are responsible for designing and implementing AI systems that balance functionality with ethical considerations [20], [21]. Their work often involves making decisions that impact user data collection and privacy protocols. For example, [21] conducted a study involving 1,015 Generation Z students across 48 countries, which demonstrated that cultural and social contexts significantly affect trust in AI. This suggests that traditional models, such as the Technological Acceptance Model, may not fully address the trust dynamics needed in cross-cultural youth interactions with AI [8]. The perspectives of AI professionals are influenced by both technological feasibility and compliance with emerging regulations [22], [23]. Research highlights that while developers acknowledge the importance of data control and user consent, their primary focus may shift toward innovation and performance [24], [25]. This divergence in priorities between educators/parents and developers underscores the need for multi-stakeholder approaches to address the privacy challenges of AI systems effectively.

TABLE I. CONSTRUCTS AND DEFINITION

| Construct | Definition |
|---|---|
| Data Ownership and Control (DOC) [26], [27], [28], [29] | It is the degree to which young people have control over their personal data and engage in discussions about privacy. |
| Parental Data Sharing (PDS) [30], [31], [32] | It is the degree to which parents exercise their rights to share children's data and consider the implications of doing so. |
| Perceived Risks and Benefits (PRB)[29], [31], [33] | It is the degree to which individuals perceive risks, ethical concerns, and benefits related to the use of personal data by AI systems. |
| Transparency and Trust (TT) [34], [35], [36] | It is the degree to which transparency in data usage influences trust in AI systems. |
| Education and Awareness (EA) [37], [38], [39], [40] | It is the degree to which stakeholders are informed about privacy and ethical issues associated with AI. |

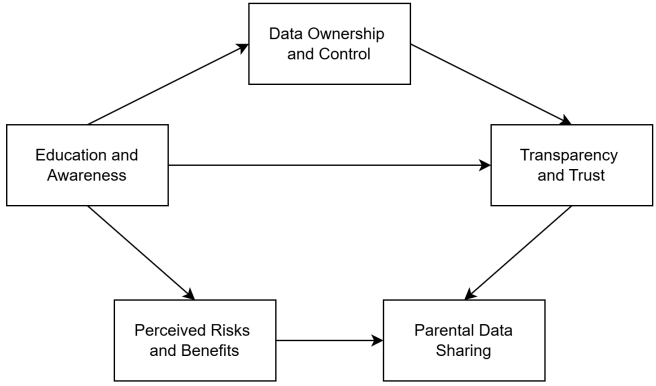

Fig. 1. Structural model for our study

## III. Methodology

In this section, we outline our research hypotheses, research questions, measurement instruments, and participant demographics.

### A. Research Model and Hypotheses:

Based on the findings from the literature review, we have developed seven research hypotheses to examine the constructs detailed in Table I, as outlined below:

- **H1:** Education and Awareness positively influence Transparency and Trust [15], [41], [42].
- **H2:** Data Ownership and Control influences Transparency and Trust [15], [19], [38], [43].
- **H3:** Education and Awareness influence Perceived Risks and Benefits [13], [14], [29], [44].
- **H4:** Transparency and Trust influence Parental Data Sharing [12], [45].
- **H5:** Perceived Risks and Benefits influence Parental Data Sharing [15], [16], [46], [47].
- **H6:** Education and Awareness influence Data Ownership and Control [13], [14], [37], [48], [49].
- **H7:** Perceived Risks and Benefits mediate the relationship between Education and Awareness and Data Sharing Attitudes [12], [13], [16], [33], [50].

### B. Research Design

The present study received ethics approval from the Vancouver Island University Research Ethics Board (VIU-REB). The approval with reference number #103116 was given for behavioral application/amendment forms, consent forms and questionnaires. We conducted a pilot study with 6 participants, including members of empirical research specialists from the University of Saskatchewan and Vancouver Island University. The pilot study aimed to assess the feasibility and duration of the research approach and refine the study design. Participants provided general feedback on the questionnaire which informed modifications and restructuring of the final survey questionnaires. The revised research model was then tested by gathering survey data.

We recruited participants through flyers, emails, personal networks, and on social networking sites, LinkedIn and Reddit. Participation was entirely voluntary and did not receive any form of compensation. The participants had to read and accept a consent form to participate in the study, by submitting the consent form before starting the questionnaire participants were indicating they understood the conditions of participation in the study outlined in the consent form. We conducted online surveys through Microsoft Forms by requesting each participant to respond to the questionnaire based on our three designated demographics: AI Researchers and Developers, Teachers and Parents, and Youth aged 16-19. The survey instruments are adapted from constructs validated in prior studies [26], [27], [28], [29], [30], [31], [32], [29], [31], [33], [34], [35], [36], [37], [38], [39], [40]. The instrument consists of 3 indicators for Data Ownership and control (DOC), 2 indicators for Parental Data Sharing (PDS), 4 indicators for Perceived Risk and Benefit (PRB), 3 indicators for Trust and Transparency (TT), 3 indicators for Education and Awareness (EA), and 3 open-ended discussion questions. The respective items (questions) within these constructs are detailed in Table II. We measured responses to the items, excluding qualitative items, on a 5-scale Likert scale. Notably, to ensure consistency in outcomes, we reversed the scale for items in PRB for AI Researchers and Developers to align contextually with the scale for items in PRB of educators and parents. The open-ended questions and 2 indicators from PRB were used for qualitative analysis, while the remaining items were used for quantitative analysis. To achieve a holistic understanding of the subject, we have also conducted a complementary qualitative study, details of which are discussed in [8].

TABLE II. CONSTRUCTS AND ITEMS

| Construct | Items |
|---|---|
| Data Ownership and Control (DOC) | doc1: Importance of users having control over their personal data.<br>doc2: Frequency of considering user data control in work.<br>doc3: Feasibility of implementing data control mechanisms. |
| Parental Data Sharing (PDS) | pds1: Handling data shared by parents on behalf of children.<br>pds2: Importance of obtaining consent from young users. |
| Perceived Risks and Benefits (PRB) | prb1: Concern about ethical implications.<br>prb2: Significance of benefits in justifying data use.<br>Qpen-Ended Question: Primary risks associated with personal data use.<br>Qpen-Ended Question: Benefits AI systems provide by using personal data. |
| Transparency and Trust (TT) | tt1: Importance of transparency about data usage.<br>tt2: Perception of transparency in current AI systems.<br>tt3: Belief that increasing transparency improves user trust. |
| Education and Awareness (EA) | ea1: Knowledge about privacy issues related to AI systems.<br>ea2: Belief that users receive adequate training on privacy.<br>ea3: Importance of being educated on privacy and ethical issues. |

TABLE III. PARTICIPANTS' DEMOGRAPHICS

| Respondents' characteristics | Percentage |
|---|---|
| Parents | 37% |
| Educators | 40 % |
| Both Parent and Educator | 23% |
| AI Developers | 34% |
| AI Researchers | 65% |
| Both AI Researcher and Developer | < 1% |

*C. Participants demographics*

Table III highlights the demographics of the participants. A total of 252 participants took part in the study: 115 were parents and/or educators, 124 were AI professionals, and 13 were young digital citizens (aged 16–19). After data cleaning, 110 valid responses from educators/parents and 117 valid responses from AI professionals remained for analysis. Young digital citizens were excluded from the current analysis due to the small number of valid responses, which was insufficient for reliable statistical analysis. Of the 110 valid educator and/or parent responses, 41 identified as parents, 44 as educators, and 25 identified as both. Among the 117 valid responses from AI professionals, 40 identified as AI developers, 76 as AI researchers, and 1 as both.

## IV. RESULTS

We used Microsoft Excel to process the data collected with descriptive statistics. We combined stakeholders, educators, parents, and AI developers into a single sample for analysis to streamline the process. We analyzed the data with a partial least squares structural equation model (PLS-SEM) using smartPLS software [51]. PLS is a well-established technique for estimating path coefficients in structural models used widely in research [52], [53]. The SEM as suggested by [54] includes the testing of measurement models (exploratory factor analysis, internal consistency, convergent validity, Dillon-Goldstein's rho) and the structural model (regression analysis). We used the path weighing structural model scheme in smartPLS which provides the highest $R^2$ values for dependent latent variables. We also applied a nonparametric bootstrapping procedure to evaluate the statistical significance of different PLS-SEM results. Bootstrapping is a resampling method that involves drawing samples with replacements from the original data to create an empirical sampling distribution. For our analysis, we generated 5,000 subsamples and performed a two-tailed test at a significance level of 0.1. We also conducted a thematic analysis of open-ended questions to identify common themes expressed by participants. This paper focuses on quantitative analysis, while the qualitative insights derived from the same research, which provide valuable context and complement the findings, are presented in a separate study [8].

*A. Descriptive Statistics*

Our quantitative survey responses were collected using a 5-scale Likert scale, enabling us to compare the mean response values across the constructs, and the difference between the two group's mean response values as shown in Fig. 2. Most constructs showed similar mean values between the responses of parents/educators and AI developers/researchers. The Data Ownership and Control (DOC) construct had a mean response value of 3.77 for educators/parents and 3.93 for researchers, signaling a high importance placed on DOC, with similar views between both groups. Similar results were observed in Transparency and Trust (TT), with a mean response of 3.45 for educators/parents and 3.46 for AI developers/researchers. Notably, the mean response value in the Data Sharing Attitude (PDS) construct was much lower, with a parent/educator mean of 2.85, and a researcher mean of 2.34. While both groups had a lower mean response in this construct, it is the only construct in which researchers had a lower mean response compared to parents. The Perceived Risks and Benefits (PRB) construct saw the highest mean responses amongst both groups, with parents/educators averaging 3.81, and researchers at 4.39, indicating a particularly high importance placed on the PRB construct in both groups. Interestingly, the Education and Awareness (EA) construct had the greatest discrepancy between parents/educators and researchers. Parents/educators had a mean response of 3.41, and researchers had a mean response of 4.12. While both groups had a relatively high mean response, it is indicated that the EA construct holds greater importance amongst the researcher group.

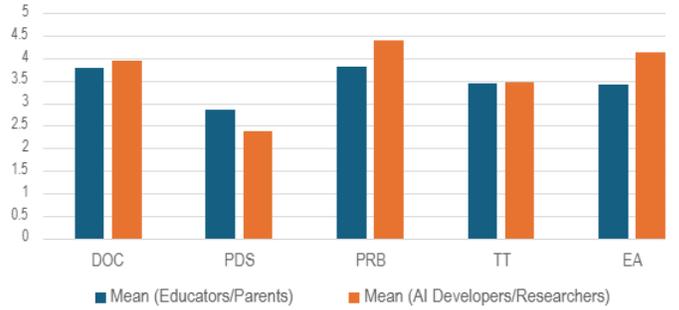

Fig. 2. Analysis of all the constructs

*B. Measurement Models*

We checked the measurement model with exploratory factor analysis by testing the internal data consistency, reliability and validity of the constructs.

*1) Exploratory Factor Analysis:* For exploratory factor analysis, we first checked the factor loadings of individual items shown in Table IV, to see how each variable loaded on its own construct over the other respective constructs. Factor loadings greater than 0.60 can be considered as significant according to [55]. In our study, all the indicators in the measurement model had a factor loading of value greater than 0.60 except for item 2 in the contrust Trust and Transparency (TT), and item 1 in the Data Ownership and Control (DOC) construct. Item tt2 had a low loading value of 0.396 which would suggest that it be avoided in the model. Although we did use the validated constructs, our exploratory analysis showed that tt2 had a weak influence on Trust and Transparency. The item doc1 had a factor loading value of 0.584, which is just under the significant level of 0.60, which is still deemed moderately acceptable [55].

TABLE IV. EXPLORATORY FACTOR ANALYSIS

| Construct | Item | Factor Loading |
|---|---|---|
| DOC | doc1 | 0.584 |
| | doc2 | 0.782 |
| | doc3 | 0.718 |
| PDS | pds1 | 0.836 |
| | pds2 | 0.889 |
| PRB | prb1 | 0.776 |
| | prb2 | 0.659 |
| TT | tt1 | 0.802 |
| | tt2 | 0.396 |
| | tt3 | 0.825 |
| EA | ea1 | 0.795 |
| | ea2 | 0.632 |
| | ea3 | 0.716 |

*2) Constructs reliability and validity:* We observed the convergent validity for each construct measure by calculating Average Variance Extracted (AVE) and Composite Reliability (CR) from the factor loadings (see Table V For adequate convergent validity, AVE should exceed 0.50, indicating that at least 50% of the variance in items is explained by the construct, and CR should exceed 0.75 [56]. In this study, AVE for each construct exceeded 0.50 except for Trust and Transparency (TT) and Data Ownership and Control (DOC). Similarly, CR for DOC and TT was also just below the suggested value of 0.75. For TT, the item tt2 had a low factor loading, contributing to an AVE of 0.494 and a CR of 0.729. This suggests that tt2 contributes less to the variance explained by the construct but may capture a unique and theoretically significant aspect of TT, emphasizing the need for its theoretical and contextual significance despite the lower loading. DOC also showed moderate factor loadings, with an AVE of 0.489 and a CR of 0.739, indicating slightly weaker convergent validity. Perceived Risks and Benefits (PRB) reported a CR of 0.681 but had an acceptable AVE of 0.518, demonstrating some internal consistency. The remaining constructs showed acceptable CR values exceeding 0.75.

Table V also presents the rho_A (Dillon-Goldstein's rho) values, a more robust reliability measure than Cronbach's alpha in SEM [57]. DOC and Education and Awareness (EA) demonstrated moderate reliability with rho_A values of 0.509 and 0.542, respectively. Parental Data Sharing (PDS) showed good reliability with a rho_A of 0.673, while PRB exhibited poor reliability with a rho_A of 0.070, suggesting it may require reconsideration for further analysis. TT achieved a rho_A of 0.575, reflecting moderate reliability.

*C. Structural Models*

The results of our proposed model built for PLS-SEM analysis [58] are shown in Figure 3.

TABLE V. CONSTRUCT RELIABILITY AND VALIDITY

| Construct | rho_A | AVE | CR |
|---|---|---|---|
| DOC | 0.509 | 0.489 | 0.739 |
| PDS | 0.673 | 0.745 | 0.853 |
| PRB | 0.070 | 0.518 | 0.681 |
| TT | 0.575 | 0.494 | 0.729 |
| EA | 0.542 | 0.515 | 0.759 |

The model is characterized by coefficients of determination ($R^2$'s), path coefficients (β's) and corresponding P-value. $R^2$ determines the variance of a given construct explained by antecedents, β captures the strength of the relationship between the selected constructs, and P-value determines the statistical significance of the model. According to Chin's guideline [31], a path coefficient (β) should be equal to or greater than 0.2 to be considered relevant. Based on [31], [59], models are deemed statistically somewhat significant (*p) when p < 0.1, statistically quite significant (**p) when p < 0.01, and statistically highly significant (***p) when p < 0.001. Table VI shows the standardized path coefficient (β), t-statistics, and p-value for the model.

The model in Fig. 3 depicts causal relationships among Data Ownership and Control (DOC), Trust and Transparency (TT), Education and Awareness (EA), Perceived Risk and Benefit (PRB), and Parental Data Sharing (PDS). While observing the direct effects, the findings revealed a significant positive effect of EA on DOC (β = 0.410; p < 0.001) and a strong positive influence of EA on PRB (β = 0.547; p < 0.001). However, EA showed an insignificant effect on TT (β = 0.076; p > 0.05). Additionally, DOC positively affected TT (β = 0.358; p < 0.001), indicating that increased data ownership and control enhance transparency and trust in AI systems. Conversely, the relationships between PRB and PDS (β = -0.098; p > 0.05) and between TT and PDS (β = -0.102; p > 0.1) were not significant. Therefore, hypotheses H2, H3, and H6 were supported, while H1, H4, and H5 were rejected.

Moreover, EA explains 16.8% of the variance in DOC ($R^2$ = 0.168), and accounts for 29.9% of the variance in PRB ($R^2$ = 0.299). Additionally, EA and DOC explain 15.6% of the variance in TT ($R^2$ = 0.156). However, TT and PRB explain only a minimal portion of the variance in PDS, at 2.5% ($R^2$ = 0.025). Considering the indirect effects, the analysis revealed that the pathway from EA to PDS through PRB was not significant (β = -0.054; p = 0.364). This indicates that while EA may influence PRB directly, the extension of this influence to PDS through PRB does not hold statistical significance. Consequently, this suggests that PRB does not mediate the relationship between EA and PDS effectively, leading to the rejection of hypothesis H7.

## V. DISCUSSION

*A. Education and Awareness Enhance Data Control and Risk Perception*

The study reveals that Education and Awareness (EA) significantly positively impact both Data Ownership and Control (DOC) and Perceived Risk and Benefit (PRB).

TABLE VI. SEM ANALYSIS

| Structural path | Std β | T | P |
|---|---|---|---|
| DOC → TT | 0.358 | 4.716 | 0.000 |
| EA → DOC | 0.410 | 6.361 | 0.000 |
| EA → PRB | 0.547 | 9.333 | 0.000 |
| EA → TT | 0.076 | 0.892 | 0.372 |
| PRB → PDS | -0.098 | 0.884 | 0.376 |
| TT → PDS | -0.102 | 1.351 | 0.177 |
| EA → PRB → PDS | -0.054 | 0.907 | 0.364 |

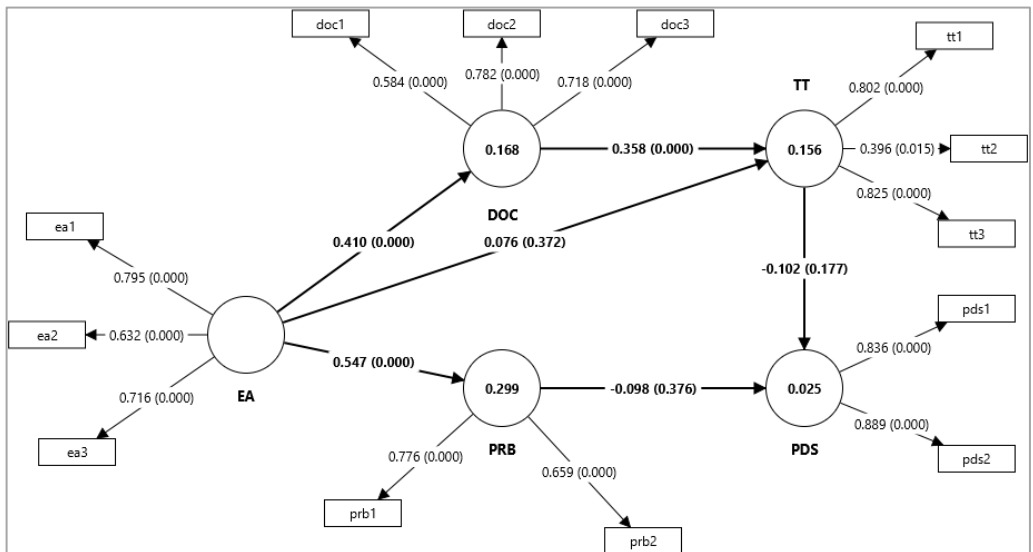

Fig. 3. Structural model showing test results (direct effect). *p < 0.1; **p < 0.01; ***p < 0.001

This suggests that when individuals are more educated and aware, they feel more empowered to control their data and have a heightened ability to assess the risks and benefits associated with data sharing. This finding aligns with privacy empowerment theories, indicating that informed users are better equipped to navigate data privacy concerns and make informed decisions regarding their personal information. These insights highlight the importance of improving educational content to make it more actionable, equipping stakeholders with practical knowledge to foster a culture of data ownership and proactive privacy management. However, the moderate reliability for DOC, slightly below the ideal threshold (AVE = 0.489, CR = 0.739), suggests that while the construct is impactful, further refinement is needed to bolster measurement consistency.

*B. Data Ownership Enhances Trust and Transparency*

The positive effect of DOC on Trust and Transparency (TT) underscores the theoretical assertion that a sense of ownership and control over personal data fosters greater trust in AI systems. When users have control over their data, they perceive the systems as more trustworthy. However, EA did not significantly affect TT directly, suggesting that while education enhances control and risk perception, it does not automatically translate into increased trust without tangible control mechanisms in place. This highlights the importance of integrating user control features to build trust in technological platforms. Also, while transparency is universally valued, the approach should vary: parents and educators benefit from simplified, clear communication, while AI professionals and tech-savvy individuals may seek deeper, algorithmic insights. The moderate reliability of the TT construct (AVE = 0.494, CR = 0.729) implies that improvements in how transparency is conceptualized and measured are necessary to fully capture its role across diverse user groups.

*C. Limited Influence on Parental Data Sharing Behaviors*

The study found that neither PRB nor TT significantly influence Parental Data Sharing (PDS). Additionally, PRB did not mediate the relationship between EA and PDS effectively. These results suggest that factors other than perceived risks, benefits, trust, or transparency may play a more significant role in parental decisions to share data. The minimal variance explained in PDS indicates that additional variables, such as cultural norms, external regulations, or perceived necessity of services, might be critical in understanding data-sharing behaviors. This calls for further theoretical exploration into the complex factors influencing parental data-sharing decisions. The lack of significant influence from PRB on PDS implies that strategies should extend beyond mitigating risks to also emphasize the potential benefits of data sharing. Balancing safety with perceived personal or societal utility may enhance user engagement and willingness to share data responsibly.

*D. Implications for Policy and AI System Design*

To address the findings, stakeholders must enhance educational programs to empower users, especially young digital citizens, to manage and protect data. Integrating privacy topics into curricula fosters early ownership. AI developers should embed user-adjustable privacy features into systems, promoting trust through transparency and control. Tailored transparency efforts are key: clear communication for parents and educators, and deeper insights for AI professionals. Strategies should balance risks and benefits, emphasizing societal and personal utility to encourage responsible data sharing. Recognizing factors like cultural norms and regulations, and collaboration among policymakers, educators, and developers is crucial for informed, secure interactions with AI technologies.

*E. Limitations*

This study has some limitations that warrant consideration. First, the study's focus on a specific population such as parents, educators, and AI professionals, may limit the generalizability of the findings to other demographic or professional groups. Next, the cross-sectional nature of the study prevents a deeper understanding of how these relationships evolve over time.

Future research could address these limitations by expanding the participant pool, incorporating longitudinal designs, and refining constructs to capture the complexity of privacy-related behaviors more comprehensively.

## VI. CONCLUSION

This study explored privacy concerns in AI systems by examining the perspectives of parents/educators and AI developers/researchers using five validated constructs: Data Ownership and Control (DOC), Parental Data Sharing (PDS), Perceived Risks and Benefits (PRB), Transparency and Trust (TT), and Education and Awareness (EA). Guided by seven hypotheses, the research adapted validated survey instruments refined and tested through a pilot study, with data collected via online surveys and analyzed using descriptive statistics and Partial Least Squares Structural Equation Modeling (PLS-SEM). Results showed EA significantly enhanced perceptions of data control and risks, while DOC positively influenced TT, emphasizing the need to empower users and build trust in AI systems. However, the limited impact of PRB and TT on PDS, and the lack of mediation by PRB, suggest broader external factors, such as cultural norms and regulations, shape data-sharing behaviors. These insights reinforced the need for user-centric privacy controls, tailored transparency strategies, and targeted educational initiatives. A thematic analysis of open-ended survey responses will be conducted to complement the quantitative findings. Future research should also refine constructs, incorporate longitudinal designs, and expand demographics to capture evolving privacy attitudes, informing AI systems that balance innovation with strong privacy protections, and fostering trust and ethical governance.

## ACKNOWLEDGMENT

This project has been funded by the Office of the Privacy Commissioner of Canada (OPC); the views expressed herein are those of the authors and do not necessarily reflect those of the OPC.